# ORÍGENES

Revista de la Asociación de Genealogía e Historia de Costa Rica

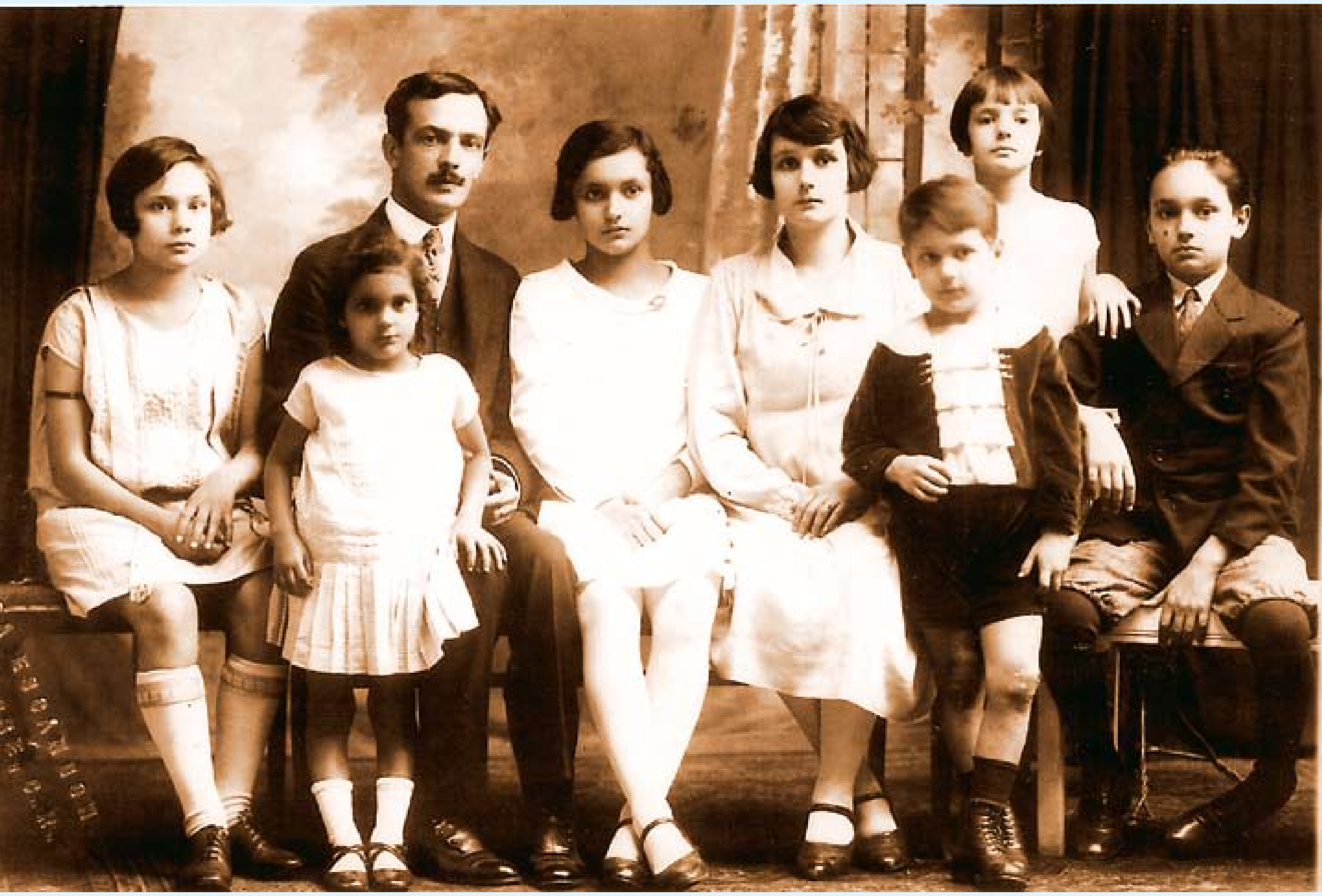

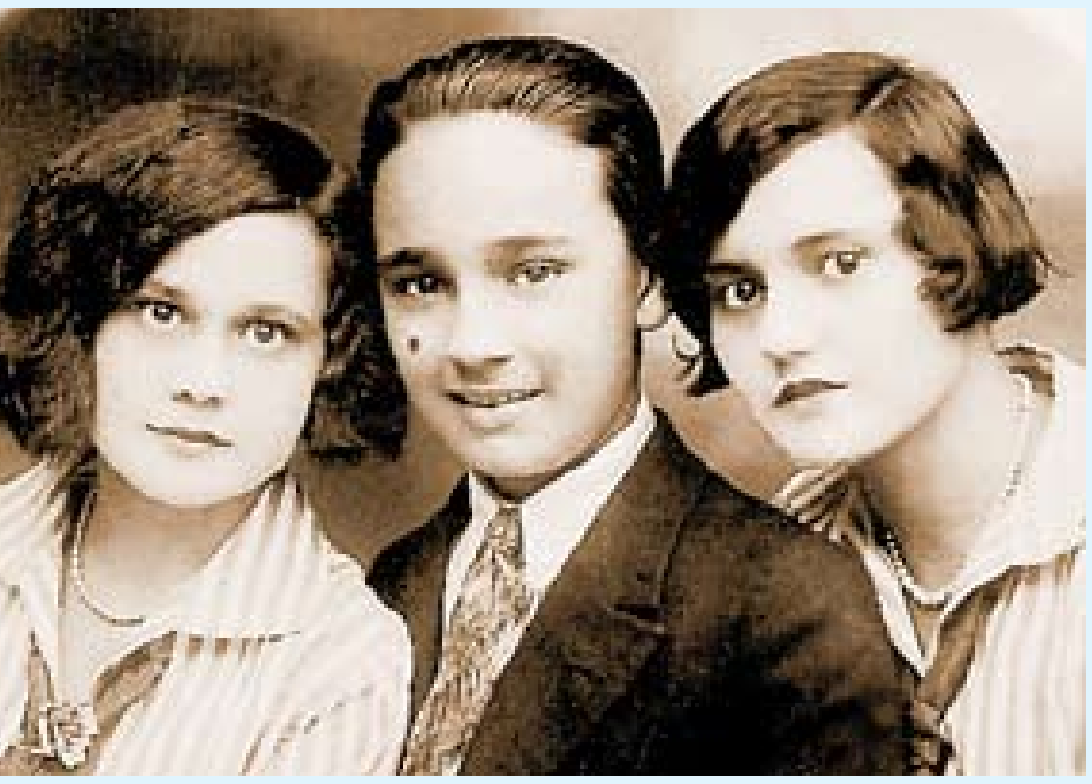

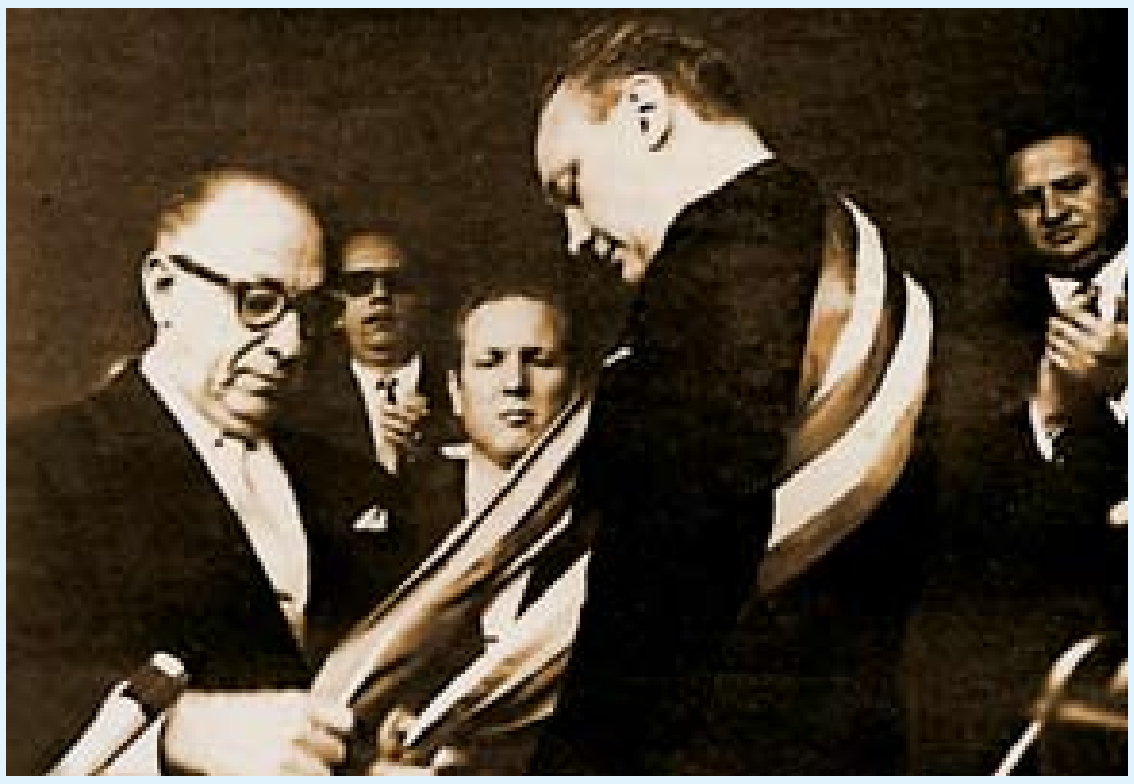

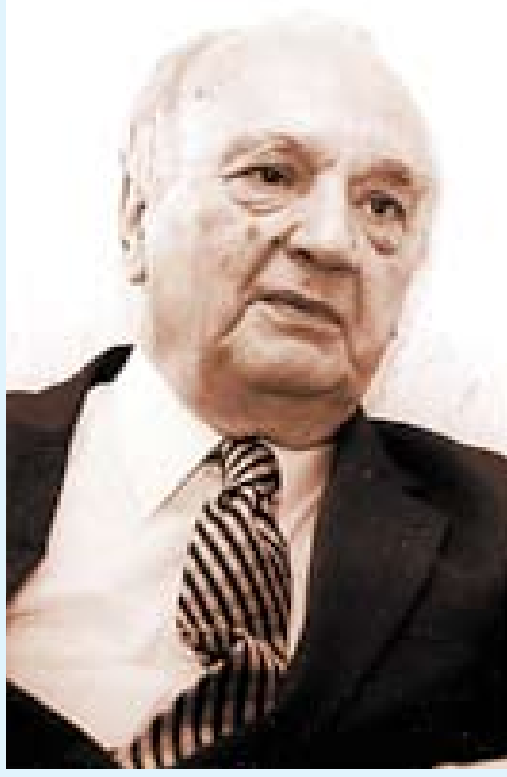

# La genealogía de doña Isabel Jiménez
## una aproximación a la primera fase del mestizaje en Costa Rica*

BERNAL MORERA BRENES **
RAMÓN VILLEGAS PALMA***
MAURICIO MELÉNDEZ OBANDO***

## I. Introducción

Hace 20 años, una niña blanca como la nieve nació en un pueblo de Costa Rica. Ella recibió un regalo inusual para un bebé, una amplia genealogía de sus ancestros. Un aspecto sobresaliente de dicha genealogía era que la línea uterina (matrilineal) pudo ser reconstruida en forma completa hasta 1580 (*véase Cuadro genealógico Nº1*), año en el que nació en Cartago doña Isabel Jiménez, importante dama de la sociedad de su época. Cartago era en aquel momento el principal centro urbano, político y administrativo en el territorio de la Provincia de Costa Rica, y hasta alrededor de 1719 fue la única ciudad española en el Valle Central; el proceso de conquista (1561-1563) recién terminaba y el dominio colonial (1563-1821) se estaba imponiendo (Acuña y Chavarría 1991).

Doña Isabel estuvo casada con don Francisco de las Alas, natural de Cartago y consignado como español. Por su parte, don Francisco era hijo y nieto de conquistadores de Costa Rica (Meléndez Chaverri 1982). Catalina de las Alas, hija de doña Isabel y su esposo, contrajo matrimonio con Pablo Milanés del Castillo, natural de Cartago e hijo de Martín Prieto del Castillo y Luisa Hernández. Ana Jiménez del Castillo (también conocida como Ana de las Alas), hija de Pablo y Catalina, casó con don Pedro Fernández de Azofaifo[1] y Tenorio, originario de Jerez de la Frontera, España, quien vino a Costa Rica con su pariente el gobernador Juan de Villalta en 1630 (Sanabria 1957, Castro Tosi 1975).

Doña Isabel, quien protocolizó su testamento en 1629, se declaró hija de don Alonso Jiménez[2]. Él había ingresado a Costa Rica con las huestes del conquistador don Juan Vázquez de Coronado –luego nombrado adelantado– (Meléndez Chaverri 1982). Siendo el testamento un documento legal que se escribía generalmente cuando se aproximaba el final de la vida y se debía enfrentar el juicio divino, los españoles usualmente no mentían en sus declaraciones. Sin embargo, en este caso particular había una omisión importante: el nombre de la madre nunca fue mencionado en el testamento, lo que trunca la posibilidad de continuar con el estudio genealógico documentado. Tal omisión es inusual, ya que lo frecuentemente observado es que durante la Colonia los hijos "ilegítimos" omitieron el nombre del padre en los documentos oficiales, pero no el de la madre. La pregunta obvia que vino a nuestra mente fue: ¿Por qué la omisión?

El papel que desempeñaron las mujeres en el establecimiento de las sociedades hispanas en el Nuevo Mundo es poco conocido. La mayoría de las veces ellas no fueron incluidas en los documentos e incluso sus nombres de pila son muy rara vez mencionados (Meléndez Chaverri 1982); motivo que le confiere un singular valor al caso de doña Isabel Jiménez.

** Laboratorio de Genética Evolutiva, Escuela de Ciencias Biológicas, Universidad Nacional, Heredia, Costa Rica.
*** Ramón Villegas Palma y Mauricio Meléndez son responsables del análisis genealógico. Bernal Morera es responsable del análisis genético y etnohistórico.

**Cuadro genealógico Nº 1**

## Genealogía matrilineal ascendente de Sofía Palma González

**Sofía Palma González**
n. SJ. 5.5.1988

**Da. Marianela González Garcia**
n. Al., 30.4.1965
casó en Atenas, Alajuela, 1987, con
Dn. Jorge Arturo Palma Murillo

**Da. María del Carmen García Campos**
n. Atenas, 18.10.1932
casó en Atenas, 1959 con
Dn. Carlos Luis González Murillo

**Da. Ester Campos Vargas**
n. Atenas, 3.10.1894
casó en Atenas, 1915, con
Dn. Abelino García Campos

**Da. Salvadora Vargas González**
n. Atenas, 17.5.1872
casó en 1888 con
Dn. José María Campos Morera

**Da. Sérvula González Solera**
n. Alajuela, 25.8.1850; sepultada en Atenas, 4.7.1913.
casó en Atenas, 1866, con
Rafael Silvestre Vargas González

**Da. Mercedes Solera Soto**
n. Alajuela, 6.4.1830; sepultada en Atenas, 27.2.1904
casó en Alajuela, 1848, con
Dn. Juan Pablo González Castro

**Da. Paula Soto Oviedo**
n. H. 9.8.1787; sepultada en Alajuela, 14.4.1856
casó en Alajuela, 1813, con
Dn. Casimiro Solera Venegas

**Da. Mercedes Oviedo Herrera**
n.h. 1759; sepultada en Alajuela, 5-4-1826
casó en Heredia, 1779, con
Dn. Isidro Soto Umaña

**Da. María Natividad Herrera Solózano**
n. H. h. 1745
casó en Heredia, 1761, con
Ignacio Oviedo González

**Da. Francisca Solórzano Morera**
m. H.20.11.1741
casó con 1735 con
Ventura Herrera Esquivel

**Da. Casilda Morera Azofeifa**
n.h. 1678
casó en 1698 con
c. Cap. Antonio Solórzano
natural de Cartagena de Indias (Colombia)

**Da. Baltasara de Azofeifa**
[tcc Baltasara Tenorio]
casó hacia 1670 con
Juan de Morera

**Ana Jiménez del Castillo**
[tcc Ana Martínez Navarro]
casó hacia 1635 con Pedro Fernández de Azofeifa
Natural de España

**Isabel Jiménez**
n.h. 1585; t.C.28.3.1629
casó 1º con Francisco de Salas
(además casó 2º con Juan López de Ortega)

**Desconocida**
n.h. 1560, tuvo descendencia con el conquistador
Alonso Jiménez
natural de Bienvenida, España

Cuadro elaborado por Bernal Morera, Ramón Villegas y Mauricio Meléndez Obando.

El conocimiento del origen étnico de las mujeres fundadoras y su contribución biológica a las modernas sociedades americanas es un aspecto relativamente difícil de rastrear desde una perspectiva documental. Se debe tener en mente que la reconstrucción genealógica de un linaje materno desde personas vivas en el presente hasta sus ancestros cuatro o cinco siglos atrás, es un asunto relativamente difícil debido a que en las sociedades hispanoamericanas los apellidos siguen los linajes masculinos en tanto que los apellidos femeninos se pierden en cada generación. Además, se requieren registros de bautismo completos y matrimonios legales en cada generación, o si faltaran, se necesita que esté disponible información de algún matrimonio consanguíneo posterior que pueda ayudar a reconstruir la filiación faltante. Otras fuentes que contienen información genealógica son los testamentos, mortuales y cartas dotes, que algunas veces son muy útiles para corroborar las genealogías, pero desafortunadamente rara vez están disponibles (Meléndez Obando 1999).

## II. ADN mitocondrial y reconstrucción de linajes maternos

Por otra parte, las tecnologías modernas de la genética molecular nos han proporcionado la posibilidad de abordar el estudio de las genealogías desde un enfoque diferente, analizando directamente los restos óseos (Gill et al. 1994, Ivanov et al. 1996, Boles et al. 1995, Lalueza-Fox et al. 2001) o estudiando a los descendientes vivos (Heyer 1995, Soodyall et al. 1997). El ADN mitocondrial (ADNmit) tiene una herencia estrictamente materna y una gran variabilidad que lo hacen útil para realizar estudios históricos y evolutivos (Stoneking 1993). Durante los últimos años, el análisis de las secuencias del ADNmit ha demostrado ser un poderoso instrumento para el estudio de la historia de las poblaciones humanas. Ha sido aplicado a muy diferentes marcos espaciales y temporales, como la colonización de continentes, el origen de poblaciones específicas y a la identificación de linajes y de individuos. En tales estudios se ha recolectado una gran cantidad de secuencias de ADNmit de poblaciones nativas, lo que ha mostrado la existencia de linajes específicos de continente: como africanos, asiáticos, europeos e indígenas americanos (Alves-Silva et al. 1999).

## III. Mezcla en Costa Rica

Los estudios recientes de antropología biológica han demostrado que la población general de Costa Rica es fundamentalmente el resultado de una mezcla trihíbrida, con proporciones de genes de origen europeo, amerindio y africano de 61,04%, 29,91% y 9,05%, respectivamente (Morera et al. 2003). Sin embargo, muchos aspectos del proceso de mezcla no se comprenden aún con claridad.

En varias otras poblaciones del continente americano que poseen una historia bien conocida de mezcla, como Brasil (Alves-Silva et al. 1999, Batista dos Santos et al. 1999), México (Merriwether et al. 1997) y Cuba (Torroni et al. 1995), el estudio de las variantes del ADNmit ha permitido encontrar diversos niveles de contribución materna indígena, africana y europea. En ausencia de documentación, estas poblaciones representan ejemplos del fenómeno de apareamiento direccional, donde los hombres europeos y las mujeres indígenas o africanas contribuyeron en forma desproporcionada a la formación de dichas poblaciones actuales. Ya están en marcha estudios cuantitativos de este proceso evolutivo en la población de Costa Rica.

En un estudio conexo, Morera et al. (2005) trazaron varias genealogías –incluida la presente– hasta las primeras generaciones de tiempos coloniales. Los resultados fueron relevantes ya que demostraron cualitativamen-te la sobreviviencia de linajes mitocondriales amerindios y europeos en la población "blanca" actual de Costa Rica.

Sin embargo, desde una perspectiva etnohistórica, estos estudios puntuales nos abren además una ventana al análisis de problemas históricos relacionados con el establecimiento de dicha población híbrida, como el origen étnico de los individuos fundadores, el aporte materno y el inicio del proceso de mestizaje. En este trabajo pretendemos hacer una aproximación más detallada de la primera fase del mestizaje en Costa Rica.

## IV. Materiales y métodos

Se construyó un árbol genealógico a partir de los datos relevantes consignados en los libros de bautismos y de matrimonios de las parroquias de Heredia, Alajuela y Atenas, cuyos originales y copias microfilmadas son custodiados en el Archivo Histórico Arquidiocesano Bernardo Augusto Thiel. Aunque en este tipo de trabajo suele omitirse el nombre de las personas vivas que han permitido realizar el estudio de su ADN, para proteger la privacidad, en este caso contamos con la autorización de las mujeres de las últimas tres generaciones descendientes de Isabel Jiménez. (*Véase Cuadro Nº 1*).

## V. Resultados

El Cuadro Nº 1 muestra la genealogía matrilineal de 17 generaciones desde doña Isabel Jiménez (1585-1629). No hay registros históricos respecto a la identidad de su madre, pero por mucho tiempo genealogistas e historiadores han asumido que su origen era español.

## VI. Discusión

### VI.1. Origen étnico documentado versus genético

A lo largo de la época colonial, los términos "don" o "doña" fueron formas de tratamiento de distinción dado a los españoles hidalgos –peninsulares y criollos–. Muchos conquistadores que no eran hidalgos ganaron lo que se llamó hidalguía de pobladuría por su participación activa en la conquista y por establecer su residencia en las ciudades recién fundadas (Castro Tosi 1975). De tal manera, normalmente, estas formas de tratamiento indican origen étnico español. Casi siempre, dichas formas de tratamiento social fueron usados en forma estricta (Meléndez Obando 1999).

Documentalmente, sabemos que Isabel Jiménez perteneció a la élite de la sociedad colonial de su época. Implícitamente, Meléndez Chaverri (1980: 151-152, 226) la considera como española. Su padre y dos de los ancestros de su esposo fueron conquistadores de Costa Rica, que recibieron honores y bienes que, según el mismo autor, los ubican dentro de los estratos superiores de las tres primeras huestes de conquista del siglo XVI (Cavallón, Vázquez de Coronado y Perafán de Rivera), o sea dentro de la aristocracia conquistadora. Sin embargo, no hemos encontrado evidencia documental de que a ella, en vida, se le atribuya el tratamiento de doña.

Con los siglos, la historia cambió, ya que Isabel Jiménez y su hija Ana han sido intituladas como "doña" en varias de las genealogías académicas de sus descendientes (Robert Luján 1955: 99, y Segura Rodríguez 1998: 67, 77, entre otros). Si bien Castro Tosi (1975) se refiere a Ana Jiménez como "una señorita costarricense, la cual si no era de primera nobleza[3], tenía muy honrados ascendientes, remontando hasta la conquista", empero, tal afirmación no sugiere ningún origen mestizo. Consecuentemente, tanto historiadores como genealogistas consideraron a Isabel como de etnia española.

Como se describió con detalle (Morera et al. 2005), el ADN mitocondrial (ADNmit) de dos individuos del linaje matrilineal de Isabel Jiménez fueron analizados y ambos presentaron la misma secuencia, con cinco diferencias respecto a la Secuencia de Referencia de Cambridge descrita por Anderson et al. (1981), en los sitios 16111T, 16223T, 16290T, 16319A y 16362C. Estas substituciones constituyen el motivo compartido por la mayoría de secuencias que pertenecen al linaje (o haplogrupo) A (Torroni et al. 1992), el cual ha sido encontrado únicamente en indígenas americanos y en asiáticos del oriente. Dicha secuencia con los cinco cambios diagnósticos ha sido previamente descrita en varias poblaciones amerindias de Centroamérica y América del Sur, tales como los ngobé (Kolman et al. 1995) y emberá de Panamá (Kolman y Bermingham 1997), en los cayapa del Ecuador (Rickards et al. 1999) y en los mapuche de Chile (Ginther et al. 1993); y no ha sido encontrada nunca en individuos de origen europeo ni africano.

A diferencia de las otras genealogías estudiadas, como la de doña Andrea Vázquez de Coronado (española) y la de Gracia (indígena) hasta costarricenses actuales (Morera et al. 2005), en este caso, sí hay una discrepancia entre la afiliación étnica biológica obtenida a partir de los resultados genéticos y la respectiva ancestría deducida a partir de la documentación histórico-genealógica, al menos en aquellos escritos del siglo XX.

En ausencia de documentos coloniales, escasos para este periodo, que pudieran proporcionar mayor información al respecto, nos inclinamos por considerar el dato genético proveniente de sus descendientes vivos como el correcto. En conclusión, Isabel Jiménez era mestiza y –considerando las fechas del proceso de conquista de Costa Rica– su madre debió ser una mujer indígena, o a lo sumo una mestiza de primera generación cuya madre (la abuela de Isabel) era ineludiblemente indígena americana.

### VII. El papel de las mujeres en la fundación de la población costarricense

Otra consecuencia de nuestro enfoque genealógico-genético es que nos permite abordar el tema del aporte materno y el inicio del proceso de mestizaje, durante el proceso de fundación de la población costarricense, en la primera fase del periodo colonial.

Dentro de este contexto, el historiador Meléndez Chaverri (1982) en su libro *Conquistadores y pobladores: orígenes histórico-sociales de los costarricenses*, opina que evidentemente las familias fundadoras dentro del Valle Central de Costa Rica, al ser el núcleo constitutivo de los estratos superiores de la sociedad global de la provincia, tendieron a amalgamarse a través del proceso matrimonial y a formar un bloque unitario, para lo cual buscaron constituir una serie de linajes, cuyo origen por lo común no iba más allá del mismo conquistador. Añade que dichas familias, fueron el punto de partida de un proceso endogámico que "quizás sea una forma de defensa para afirmar la individualidad racial del grupo español aquí asentado, elemento determinante para la distinción entre dominantes y dominados".

Meléndez Chaverri (1982) aclara que existe un tópico que él no ha analizado suficientemente, por razón del grado de dificultad que halló en la documentación que le fue posible utilizar. Referido de modo concreto a los hechos que se vinculan a la mujer y al papel por ella desempeñado en todo este proceso. Así, se mencionan unas muy pocas mujeres españolas de la generación parental que vinieron al inicio de la colonia, y otras tres esposas que se pueden excluir del análisis pues nunca vinieron o no dejaron hijos en Costa Rica.

Recientemente Meléndez Obando (2004) retoma el tema y logra rescatar el nombre de 59 esposas o "compañeras" de los 103 conquistadores o pobladores, identificados hasta la fecha, de igual forma restringido artificialmente al periodo 1561-1599 (véase Anexo N°1). Da un reducido número de mujeres de las que consta su calidad de españolas: doña Mayor de Benavides (cc Juan de Solano), doña Isabel Ruiz de Villatoro (cc Alonso del Cubillo), doña Juana Chacón (cc Francisco Pavón), doña Andrea Vázquez de Coronado (cc Diego Peláez de Berríos), doña Sabina de Artieda (cc Juan de Peñaranda), doña Tomasina de Lerma (cc Agustín Félix de Prendas), doña Marina de Anagas (cc Alonso Pérez Farfán), doña Inés de Benavides (cc Francisco de Ocampo Golfín), doña Petronila de Paz (cc Pero Afán de Ribera), doña Isabel Arias Dávila (cc Juan Vásquez de Coronado) y doña María de Céspedes (cc Diego de Artieda Chirino). De ellas, solo se sabe con certeza que Mayor de Benavides, Isabel Ruiz de Villatoro, Sabina de Artieda, Tomasina de Lerma, Marina de Anagas y María de Céspedes nacieron en España.

Y el uso de doña sugiere el mismo origen para doña Inés de Ampuero (¿cc Francisco Pavón?) y doña Ana [sic] (cc Antonio de Carvajal), pero no se cuenta con mayor información.

Meléndez Chaverri (1982) apunta que se dan unos pocos casos en los que la tolerancia inicial permitió el mestizaje, como sucedió con Antonio Álvarez Pereira, cuya descendencia ha sido relevante en Costa Rica, habida con una mujer indígena proveniente de Quepo que supuestamente tomó el nombre español de doña Inés[4], caso que adquiere mayor notoriedad al tratarse de la figura relevante de un capitán; o el caso de uno de los hijos de Juan López de Ortega, habido con una mujer indígena (con nombre españolizado María López). Meléndez Obando (2004) adiciona los nombres de tres indias: Gracia (tuvo sucesión con Domingo Jiménez), Catalina Tuia (tuvo sucesión con Francisco de Ocampo Golfín), Inés (cc Juan Pérez) y dos consideradas tradicionalmente mestizas: Francisca Gutiérrez de Sibaja (cc Diego de Quesada) y doña Inés Álvarez Pereira[5] (cc Bartolomé Sánchez) hija esta última de la citada doña Inés, lo cual sugiere un origen americano para siete de las mujeres fundadoras. Esos pocos ejemplos, a juicio de Meléndez Chaverri (1982, p. 153-154), muestran la efectiva apertura que hubo en el proceso inicial de la conquista, para tratar de fortalecer el acercamiento entre la sociedad conquistadora y la conquistada. "Pero estos vienen en cierto modo a ser, al menos documentalmente, bastante esporádicos si los comparamos con el proceso en su conjunto." De todas las demás, realmente se desconoce su origen étnico.

Por otra parte, nuestros resultados genético-genealógicos, restringidos no tanto a la disponibilidad de documentos coloniales como a la existencia de descendientes vivos por línea estrictamente materna, clarifican que la mezcla español-indígena comenzó desde las primeras generaciones de la sociedad hispana (Morera et al. 2005). Lo cual viene a corroborar la afirmación casi intuitiva de Meléndez Chaverri (1982: 157) sobre el "modo en que se inicia además el proceso de mestizaje, que es claro que se realizó, aun cuando nos sea difícil aportar pruebas (documentales) concluyentes, concretas y específicas." Así, los datos biológicos muestran que debió haber ocurrido un cierto flujo de genes amerindios hacia el grupo español dominante[6] durante las primeras generaciones de la sociedad colonial, en contraste con lo que comúnmente se ha afirmado, de que la élite española evitaba las uniones exogámicas con otros grupos étnicos. Además, esto sugiere que la contribución materna indígena a la población de Costa Rica debió ser más importante de lo que usualmente se piensa (Morera et al. 2005).

## VIII. Memoricidio

En muchas partes de Latinoamérica cuando se habla de las raíces etnohistóricas, la gente se remite, sistemáticamente, a Europa, y olvida el origen pluriétnico de nuestros pueblos. La razón de ese "olvido" tiene su fundamento principal en el racismo que se proyectaba contra el indígena y el negro, quienes fueron colocados en la base de la pirámide social durante la Colonia. Por eso resultó sencillo, para muchísimas familias americanas, borrar de su memoria la parte de su origen que consideraron oscura, indeseable, y exaltaron aquella que las vinculaba a España. En Costa Rica, el olvido de estas raíces pluriétnicas ha sido una constante, promovida por genealogistas preocupados –exclusivamente– por hallar "hidalguías" o "purezas de sangre" y por una historiografía oficial que soslaya el origen diverso, abstrayendo a nuestra nación de la realidad latinoamericana (Meléndez Obando 1996, 2000).

Y es en este sentido en el que el caso de doña Isabel Jiménez adquiere especial relevancia: ¿Fue la omisión del nombre y origen de su madre en el testamento un hecho casual o deliberado? En nuestra opinión, ella omitió intencionalmente el nombre de su madre porque se identificaba con la etnia conquistadora y para que no quedara evidencia documental de sus orígenes indígenas. Un poco más adelante en la Colonia, los hijos "ilegítimos" –como la misma Isabel– contrariamente omitían de sus testamentos el nombre del padre, para no evidenciar documentalmente su "pecado", aun cuando utilizaban su apellido y probablemente todos sus contemporáneos conocían quién era. Lo que resulta llamativo es que el "memoricidio" de las raíces indígenas haya surgido desde la primera generación mestiza, durante un periodo en el que la tolerancia inicial permitió e incluso facilitó el mestizaje.

## X. El mito de la Costa Rica "blanca" y "homogénea"

Uno de los mitos más consolidados de la historiografía tradicional y de la mentalidad popular en Costa Rica es la "españolidad" de sus habitantes, por lo que, en la actualidad la mayoría de los costarricenses se consideran a sí mismos como "blancos", claramente diferenciados por su apariencia fenotípica del conjunto centroamericano (Meléndez Obando 1996, Acuña-Ortega 2001). Además –como se comentó– la elite político-económica ha mantenido el conocimiento genealógico como garantía de la afiliación a su clase, como descendientes directos de la "dinastía de los conquistadores" (Stone 1982). Como hemos afirmado antes (Morera et al. 2003), la idea de la supuesta "pureza" de la población de Costa Rica, fue promulgada originalmente por los políticos e intelectuales liberales en el siglo XIX (Molina 1849, Biolley 1892), cuando estaban tratando de consolidar una identidad nacional diferente y separada del resto de Centroamérica, y promover la inmigración europea. Este concepto erróneo ha sobrevivido por dos siglos, proyectándose incluso a través de los grandes cambios ideológicos en la política nacional. Así, uno de los principales arquitectos del estado socialdemócrata costarricense sintetizó la idea al describir a Costa Rica como un país con una "composición étnica blanca y homogénea" (Facio 1942). Como consecuencia, este mito ha producido un pueblo ignorante de sus orígenes mestizos y que se siente superior a sus vecinos centroamericanos, a quienes llama despectivamente "indios". (Meléndez Obando 1996).

Por el contrario, los resultados biológicos y etnohistóricos recientes no apoyan la idea de la supuesta "españolidad", "pureza" u "homogeneidad" genética del pueblo de Costa Rica, ni de la población del Valle Central, sino que indican una situación compleja y mucho más cercana a la realidad latinoamericana (Morera y Barrantes 1995, Barrantes y Morera 1999, Morera et al. 2003, 2005, 2010).

## Agradecimientos

Esta investigación fue apoyada por GICYT (España) concesiones PB95-0267-C02-01 y PB98-1064 y por la Generalitat de Catalunya, Grup de Recerca Consolidat 2000SGR00093. Morera Brenes recibió una beca doctoral del AECI (Agencia Española de Cooperación Internacional).

## BIBLIOGRAFÍA

## NOTAS

[1] **Nota del editor:** Este apellido luego se transformó en Azofeifa.

[2] **Ficha biográfica:** Alonso Jiménez nació en 1544 en Bienvenida, Extremadura, España. Salió hacia América el 19 de setiembre de 1565, en el viaje que planificó desde España Juan Vázquez de Coronado y en el que este último pereció. Arribó a la provincia de Costa Rica alrededor de 1566. Fue con Perafán de Ribera a la jornada del río de La Estrella y más tarde figuró como uno de los fundadores de la efímera ciudad de Nombre de Jesús, donde fue regidor del primer cabildo de la ciudad. Luego fue alcalde ordinario y de la Santa Hermandad en Cartago. Perafán de Ribera lo favoreció en 1569 con la encomienda de Aoyaque con 300 indios, y en 1592 era encomendero en Atirro. Figuró en el gobierno de Anguciana de Gamboa, fue en tiempo de Artieda Chirinos, a una entrada a Chirripó y Aoyaque para castigar a los indígenas que se hallaban sublevados y habían matado algunos españoles. Entre 1594 y 1595, y en 1602, figuró como sargento mayor de la provincia. En 1607 fue regidor de la ciudad de Cartago. En 1600 fue tesorero provincial y se le cita aún vivo en 1610. Murió alrededor de 1610. Descendencia: Juan e Isabel (Meléndez Chaverri 1982, p. 226 y 254). Existe una biografía suya (Víquez Segrega 1959).

Alonso Jiménez, soltero, hijo de Hernán Mateos y Juana Martín, natural de la ciudad de Bienvenida, España, recibió licencia para viajar hacia Costa Rica, el 19 de setiembre de 1565. Romero Iruela, Luis y María del Carmen Galbis Díez. *Catálogo de pasajeros a Indias*. Vol. IV, 1560-1566. Pág. 513. Juan Jiménez, natural de Fuente de Cantos, soltero, hijo de Hernán Mateos y Juana Martín, (por tanto hermano carnal de Alonso), recibió licencia para viajar a Costa Rica, el 5 de setiembre de 1565, al servicio de fray Lorenzo de Bienvenida y de 13 religiosos más. Romero Iruela, Luis y María del Carmen Galbis Díez. *Catálogo de pasajeros a Indias*. Vol. IV,1560-1566. Pág. 491.

[3] El concepto de "primera nobleza" aquí utilizado por Castro Tosi (1975) es muy cuestionable. En Costa Rica nunca hubo una "alta nobleza". Tanto el título de "Adelantado" de los Vázquez de Coronado, como los pretendidos "señoríos" e "hidalgías" con los que algunos académicos actuales quieren honrar a sus ancestros coloniales, no pueden ser considerados más que como "baja nobleza". Por lo tanto, la distinción entre "primera nobleza" y "no de primera nobleza" puede resultar ilusoria.

[4] Realmente no existen evidencias documentales de que haya existido un matrimonio entre el Cap. Antonio Álvarez Pereira y Dulcehe, llamada por algunos genealogistas "princesa de Quepos". No hay, pues, evidencias de que Dulcehe y doña Inés sean la misma persona (Meléndez Obando, 2000). Algunos genealogistas ven en el "don" y "doña" que reciben los hijos del capitán Pereira la posibilidad de que los tuviera con una "india principal" o de "rango regio", pero igual se podría pensar que su madre fue una "señora principal" española cuyo nombre completo no se citó en el único documento que llegó hasta nuestros días; mismo caso que el de "doña Ana" [sic] esposa de Antonio de Carvajal, conquistador.

[5] En el Cuadro Nº2, Meléndez Obando pone en duda que doña Inés Álvarez Pereira haya sido mestiza (véase nota 4). Según la versión tradicional, doña Inés y su hija doña Inés Álvarez Pereira habrían sido consignadas como "doñas" pues pertenecían a la nobleza indígena de Quepos, aliada de los españoles.

[6] Estamos ante el grupo que se convirtió en élite de la Cartago colonial; todos culturalmente españoles, pero algunos de ellos con cierto grado de mezcla biológica con indios.

## ANEXO Nº1
**Conquistadores y pobladores (1561-1599)**

Conquistador o poblador,
con su esposa o "compañera"
1. Álvaro de Acuña (1535) esp VC
Catalina de Acuña e
2. Pero Afán de Ribera* (1492) esp
Da. Petronila de Paz e
3. Diego de Aguilar (1548) VR
Catalina Palacios e
4. Hernando de Aguilar (1545) ART
5. Pero Alonso de las Alas (1536) C
María de Guido nic e hc
6. Cristóbal de Alfaro (1540) P
Catalina Gutiérrez Jaramillo e hc
7. Juan Alonso** (1534) C
8. Antonio Álvarez Pereira (1530) por C
Da. Inés [¿india?***]
9. Francisco de Arrieta (1575) F
Catalina Gómez e
10. Diego de Artieda Chirinos* esp
María de Céspedes e
11. Juan Barbosa C
María Verdugo nic e ¿hc?
12. Martín Beleño** F
Ana de los Ríos e esp
13. Román Benito (1530) C
Juana Gómez e
14. Alonso de Bonilla (1556) nic VC
Ana López de Ortega e hc
15. Francisco de Bonilla** VC
16. Ambrosio de Brenes (1569) F
María de Espinosa e hc
17. Jerónimo Busto de Villegas C
18. Juan Cabral** ART?
Catalina Gutiérrez Jaramillo e hc
19. Miguel Calvo esp F
Mariana Chinchilla hond e
20. Antonio de Carvajal VC
Da. Ana e
21. Diego del Casar Escalante P
22. Luis Cascante de Rojas F
Juana Solano ¿mst? e hc
23. Ignacio Cota (1530) esp C
24. Alonso del Cubillo (1540) esp A
Da. Isabel Ruiz de Villatoro esp e
25. Cristóbal de Chaves (1569) F
María de Alfaro e hc
26. Gaspar de Chinchilla (1540) F
Catalina Palacios e hc
27. Gaspar Delgado (1540) P
María del Castillo e hc
28. Pedro Díaz de Loría (1538) VC
Magdalena Ojeda e
29. Luis Díaz de Trejo** C
30. Pedro Enríquez de Cadórniga** esp P
31. Francisco de Estrada** (1527) C
32. Luis de Esquivel Añasco F
33. Alonso Fajardo** VC
34. Hernando Farfán C
Antonia de Trujillo e y Catalina
Rueda e nc
35. Jerónimo Felipe (1568) esp F
María de Ortega e hc
36. Leandro de Figueroa (1566) esp F
Inés Solano ¿mst? e hc
37. Pedro de Flores (1554) A
Isabel Juárez e hc
38. Francisco de Fonseca VC
Catalina Hernández e
39. Juan García P
Juana Carrillo e
40. Pedro García Carrasco (1542) VC
Jerónima de Ávila e
41. Francisco Ginovés [Ferreto] (1512) C
42. Alonso Gómez Macotela F
43. Baltasar González (1532)** VC
44. Alonso de Guido (1543)
Isabel Núñez e
45. Alonso Gutiérrez de Sibaja (1541) guaC
María Álvarez de Oviedo nic e

46. Domingo Hernández (1536) C
47. Antonio Hernández Camelo VC
48. Antonio de Herrera** VC
49. Francisco Hidalgo** A
    Inés Pérez e
50. Alonso Jáimez esp F
51. Diego Jáimez** F
52. Gómez Jaramillo P
    Magdalena Gutiérrez e
53. Alonso Jiménez (1544) esp VC
54. Domingo Jiménez (1534)
    VR Gracia [india]
55. Juan Jiménez** esp VC
56. Francisco Lobo de Gamaza**
    (1534) VC
57. Fernando López de Azcuña F
58. Juan López Cerrato de Sotomayor****
    esp A
59. Juan López de Ortega (1546) P
    Catalina de Ortega e y María
    López [india]
60. Sebastián López de Quesada P
61. Cristóbal de Madrigal VC
    Elvira Gómez e
62. Francisco Magariño VC
63. Pedro Luis Medina Cueto F
    Isabel de Carvajal e hc
64. Esteban de Mena (1532) C
65. Vicencio Milanés (1522)** it
66. Martín de Miranda** VC
    Inés Hidalgo e
67. Felipe Monge (1565) esp F
    Francisca López e hc nc
68. Morales F
69. Francisco de Ocampo Golfín (1570) esp
    F Inés de Benavides e hc y
    Catalina Tuia [india]
70. Francisco Ochoa (1540) ART
71. Juan Ordóñez del Castillo (1535) C
72. Matías de Palacios (1550) P
    Luisa Hernández e hc
73. Francisco Pavón** ART
    ¿Da. Inés de Ampuero e?,
    Da. Juana Chacón e
74. Diego Peláez de Berríos (1565) F
    Da. Andrea Vásquez de Coronado e nc
75. Juan de Peñaranda (1535) esp ART
    Sabina de Artieda esp e
76. Antonio de Peralta esp VC
    Juana del Moral e
77. Gaspar Pereira Cardoso (1570) por F
    Isabel de Acuña e hc
78. Juan Pereira VC?
79. Juan Pérez (1522) F
    Francisca de Ávila e Inés [india]
80. Alonso Pérez Farfán (1527) esp C
    Marina de Anangas esp [no vino a CR]
81. Pedro de la Portilla (1547) A
    Ana Gómez e hc
82. Agustín Félix de Prendas** (1558) ART
    Beatriz Fernández y Da. Tomasina
    de Lerma
83. Diego de Quesada (1552) esp ART
    Francisca Gutiérrez de Sibaja mst e hc
84. Diego Quintero** C
85. Francisco Ramiro Corajo P
    Francisca de Zúñiga e
86. Jerónimo de Retes (1560) esp F
    María de Ortega e hc
87. Pedro de Ribero y Escobar (1538) VC
    Catalina de Vega e
88. Nicolás Rodas (1567)** F
89. Francisco Rodríguez** P
90. Gaspar Rodríguez (1550) A
    Inés Rodríguez e y María Ramírez e
91. Juan Rodríguez Calderón** (1550) P
92. Diego Rodríguez Chacón VC
93. Domingo Rodríguez Portugués F
94. Bartolomé Sánchez (1555) P
    Da. Inés Álvarez Pereira ¿mst? e hc
95. Miguel Sánchez de Guido (1528) esp C
    Leonor de Mendoza e hc
96. Juan Solano (1538) esp C
    Da. Mayor de Benavides esp e
97. Juan Suazo*****

Isabel Rodríguez e
98. Salvador de Torres (1576) esp F
Inés Pérez Farfán e hc
99. Diego de Trejo** (1538) C
100. Jerónimo Vanegas (1551) mex VR
Teresa Fernández [no vino a CR] e
101. Juan Vázquez de Coronado esp C
Da. Isabel Arias Dávila [no vino a CR] esp e
102. Blas de Vera Bustamente** VC
103. Diego de Zúñiga F

## NOTA DEL ANEXO

Se consideran fundadores aquellos que ingresaron antes de 1599 y dejaron descendencia que vivía aún en la primera cuarta parte del siglo XVII. Como hemos seguido el modelo de Meléndez Chaverri (base de este listado), habrá un sesgo pues él no considera a las mujeres como fundadoras, solo bajo el concepto grupal de familia (aunque no necesariamente la hayan integrado como tal bajo la figura del matrimonio). Asimismo, si se considerara fundador a cada español que arribó en ese periodo, la cifra subiría considerablemente pues algunos llegaron con hijos varones y mujeres, como Alonso del Cubillo y Juan de Peñaranda, cuyos hijos no están en la presente lista.

(*) Aunque Meléndez Chaverri los incluye como fundadores, no los consideramos tales pues no hay prueba documental, ni indicios, de que hayan dejado sucesión en Costa Rica.

(**) Caso dudoso porque no hay pruebas contundentes de que hayan dejado descendencia.

(***) Aunque se ha supuesto siempre que doña Inés fue india, en realidad nadie ha aportado prueba documental alguna; el hecho de que se le cita solo como doña Inés, no implica necesariamente que se trate de una india principal, quizá simplemente estamos ante la omisión de su apellido; lo mismo ocurrió con la esposa de Antonio de Carvajal, citada como doña Ana (sin apellido tampoco).

(****) La filiación dada por Meléndez Chaverri es totalmente equivocada. Juan López Cerrato de Sotomayor fue hijo legítimo de Alonso Fernández de Córdoba Sotomayor y Da. Inés Cerrato [esta última hija legítima del Dr. Juan López Cerrato y María de Contreras, naturales de España, quienes también tuvieron un hijo nombrado Juan López Cerrato, vecino de Granada, Nicaragua]. Tampoco lo consideramos fundador.

*****Se ignora cuándo ingreso a Costa Rica, pero sabemos que fue antes de 1599. Se le cita como poblador antiguo de la provincia.

**Abreviaturas:** A: entró con Anguciana de Gamboa; ART: entró con Artieda Chirinos; C: entró con Cavallón; CR: Costa Rica; Da.: doña; e: esposa; esp: natural de España; F: entró entre 1590 y 1599; gua: natural de Guatemala; hc: hija de conquistador; hond: Honduras; mex: natural de México; mst: mestiza; nc: nieta de conquistador; nic: natural de Nicaragua; P: entró con Perafán de Ribera; por: portugués; VC: entró con Vásquez de Coronado; VR: entró con Venegas de los Ríos.

FUENTE: MELÉNDEZ CHAVERRI (1982) Y ACTUALIZACIÓN DE MELÉNDEZ OBANDO (2004).